\documentclass[a4paper]{jpconf}
\usepackage{graphicx,url,amssymb}

\newcommand{\td}[3]{\frac{d^{#3} #1}{d {#2}^{#3}}} %total derivative of #1 w.r.t #2 of order #3
\renewcommand{\v}[1]{\ensuremath{\mathbf{#1}}} % for vectors

\renewcommand{\bar}[1]{\ensuremath{\overline{#1}}}
\begin{document}
\title{Exploring the Potential of the Dark Matter Candidate from the Madala Hypothesis with Multi-frequency Indirect Detection}

\author{Geoff Beck \& Sergio Colafrancesco}

\address{School of Physics, University of the Witwatersrand, Private Bag 3, WITS-2050, Johannesburg, South Africa}

\ead{geoffrey.beck@wits.ac.za}

%Idea: we study the fraction of DM provided by Madala related particles assuming Higgs-like couplings to SM by scalar mediator. We calculate 3sigma cross-section limits and use these to infer whether Madala DM is a sub-dominant contributor to cosmological abundance or not.

\begin{abstract}
The Madala hypothesis was proposed by members of the Wits-ATLAS group to account for several anomalies in both ATLAS and CMS data at the LHC. This hypothesis extends the standard model through the addition of two scalar bosons and a hidden sector that can provide a dark matter candidate. This hidden sector interacts with the standard model only through the mediation of one of these scalars $S$. The couplings of $S$ are not amenable to investigation in current collider data and so are assumed to be Higgs-like to reduce the parameter space of the model. Our previous work~\cite{gsmadala} has shown that these couplings can be limited via indirect dark matter detection experiments in gamma-rays (for resonant annihilations into $S$). Here we will treat the dark matter and $S$ masses independently, and we generalise our previous work~\cite{gsmadala} and examine what fraction of the cosmological dark matter abundance can be accounted for by particles in the hidden sector of the Madala hypothesis dark matter when these annihilate to standard model products via a Higgs-like $S$. We will also extend our gamma-ray analysis of Madala hypothesis dark matter to include the constraints of diffuse radio data from the Coma galaxy cluster in addition to the Fermi-LAT gamma-ray data from both this target and the Reticulum II dwarf galaxy.

Our analysis indicates that either the Madala hypothesis cannot provide the bulk of cosmologically relevant dark matter, or the $S$ boson cannot be simply Higgs-like. These apply unless the candidate particle exceeds a mass of $\sim 200$ GeV. Both these scenarios may reduce the attractiveness of the hypothesis as the second case will imply that many free parameters must be added to describe $S$, greatly weakening fit significances for the model. To investigate the full consequences of this further work will necessitate using larger astrophysical data sets to strongly constrain details about $S$.
\end{abstract}

\section{Introduction}

The Madala hypothesis has three important constituents: a heavy Higgs-like Madala boson $H$, a mediator scalar $S$, and a Dark Matter (DM) candidate $\chi$. This hypothesis was put forward to explain anomalies seen in both ATLAS~\cite{atlas-docs} and CMS~\cite{cms-docs}, particularly in the transverse momentum of the Higgs boson as well as event excesses in multi-lepton final states~\cite{madala1,madala2,madala3,madala4}. The scalar mediator $S$ is introduced to mitigate problems in quartic couplings~\cite{madala2}, through it, $\chi$ can interact both with $H$ and the Standard Model (SM). Given that the run-1, preliminary run-2~\cite{madala3} and the latest run-2 releases~\cite{cern1}, data from the Large Hadron Collider (LHC) did not remove the excesses attributed to the Madala particles, and that it provides a candidate for the missing content of the universe, it is worthwhile to examine the properties of the model from an astrophysical standpoint.\\
Here we can investigate the $\chi$ properties and thus those of $S$, which cannot be strongly limited by current collider data~\cite{madala2}, through methods of indirect DM detection. That is, we can place limits on the properties of pathways from $\chi$ to the SM by predicting resulting fluxes of gamma-rays, synchrotron and Inverse Compton (IC) emission within cosmic structures and comparing these to known spectra/upper-limits for these target environments.

In this work we will use diffuse radio data from the Coma galaxy cluster~\cite{coma-radio2003} as well as Fermi-LAT~\cite{fermi-docs} gamma-ray limits on both Coma~\cite{fermicoma2015} and the Reticulum II dwarf galaxy~\cite{Fermidwarves2015} to examine the consequences of the simplifying assumptions used to describe $S$. This being that $S$ has Higgs-like couplings to the SM. In order to model this we will use decay branching data for Higgs-like particles from~\cite{smhiggs}. We will do this by determining $3\sigma$ confidence level upper-limits on the $\chi\chi \to $ SM annihilation cross-section and comparing these to the canonical relic values~\cite{steigmann}. If the derived limits rule out the allowed relic values, then it is sufficient to say that $\chi$ cannot constitute all cosmologically relevant DM. We then take the ratio of the derived cross-section with lower-limit of the relic band to determine the maximal allowed fraction of the cosmological abundance of DM that can be composed of $\chi$ particles from the Madala hypothesis. Unlike our previous work~\cite{gsmadala}, where we study whether astrophysical data allow for Higgs-like $S$, we do not study only the resonant case $m_S = 2 m_{\chi}$, we allow $m_S$ and $m_\chi$ to be independent. We will also extend the work from \cite{gsmadala} by determining upper-limits on the branching ratio of $S$ to $W$ bosons if $\chi$ is assumed to constitute all of DM by making use this generalised mass scenario and by including radio data. We choose the $W$ boson channel as it was the most promising in previous analysis~\cite{gsmadala}.

We find that, with both Coma radio and Reticulum II gamma-ray data, we can limit $\chi$ particle DM with a Higgs-like $S$ mediator to $\mathcal{O}$(10\%) of the cosmological DM abundance, and can rule out Higgs-like couplings to $W$ bosons for a broad range of $\chi$ masses, $\sim$ $10$ - $250$ GeV, using both radio and gamma-ray data.

This paper is structured as follows: in Section~\ref{sec:ann} we explain our DM annihilation models with resultant emissions and halo details discussed in \ref{sec:emm}. The results are shown and discussed in Section~\ref{sec:res}.

\section{Dark Matter Annihilation}
\label{sec:ann}

The Madala DM $\chi$ particles annihilate to $S$ bosons that can decay to SM particles~\cite{madala2}. The annihilation cross-section found by astrophysical probes will be an effective one, from $\chi\chi \to$ SM.

The source function for particle $i$ (electrons/positrons or photons) with energy $E$ from a $\chi\chi$ annihilation and subsequent $S$ decay is taken to be
\begin{equation}
Q_i (r,E) = \langle \sigma V\rangle \sum\limits_{f}^{} \td{N^f_i}{E}{} B_f \left(\frac{\rho_{\chi}(r)}{m_{\chi}}\right)^2 \; ,
\end{equation}
where $r$ is distance from the halo centre, $\langle \sigma V \rangle$ is the non-relativistic velocity-averaged annihilation cross-section, $f$ labels the annihilation channel intermediate state with a branching fraction $B_f$ and differential $i$-particle yield $\td{N^f_i}{E}{}$, $\rho_{\chi}(r)$ is the radial density profile of $\chi$ particles in the halo, and $m_{\chi}$ is the $\chi$ mass. The $f$ channels used will be quarks $q\bar{q}$, electron-positron $e^+e^-$, muons $\mu^+ \mu^-$, $\tau$-leptons $\tau^+\tau^-$, $W$ bosons $W^+W^-$, $Z$ bosons $ZZ$, and photons $\gamma\gamma$.

The yield functions $\td{N^f_i}{E}{}$ are taken from the Pythia routines in DarkSUSY~\cite{pythia,darkSUSY} as well as \cite{ppdmcb1,ppdmcb2} and the model independent formulation within the micrOMEGAs package~\cite{micromegas1,micromegas2}.

\section{Indirect Detection in Coma and Reticulum II}
\label{sec:emm}

For the DM-induced $\gamma$-ray production, the resulting flux calculation takes the form
\begin{equation}
S_{\gamma} (\nu,z) = \int_0^r d^3r^{\prime} \, \frac{Q_{\gamma}(\nu,z,r)}{4\pi D_L^2} \; ,
\end{equation}
with $Q_{\gamma}(\nu,z,r)$ being the source function for frequency $\nu$ and position $r$ within the given DM halo at redshift $z$, and $D_L$ is the luminosity distance to the halo.
In the case of Reticulum II, we will instead calculate the resulting gamma-ray flux based on the astrophysical J-factor:
\begin{equation}
J (\Delta \Omega, l) = \int_{\Delta \Omega}\int_{l} \rho^2 (\v{r}) dl^{\prime}d\Omega^{\prime} \; , \label{eq:jfactor}
\end{equation}
with $\rho (r)$ being the halo density profile, the integral being extended over the line of sight $l$, and $\Delta \Omega$ is the observed solid angle.
The flux is then found to be
\begin{equation}
S_{\gamma} (\nu,z) = \langle \sigma V\rangle \sum\limits_{f}^{} \td{N^f_i}{E}{} B_f J(\Delta \Omega,l) \; .
\end{equation}
The halo of Reticulum II is found to have $J = 2\times 10^{19}$ GeV$^2$ cm$^{-5}$~\cite{bonnivard2015}.

For the Coma galaxy cluster, the local emissivity for the $i-th$ emission mechanism (synchrotron, ICS, bremsstrahlung) can then be found as a function of the electron and positron equilibrium distributions as well as the associated power (for power functions $P_i$ see \cite{gs2016,Colafrancesco2006})
\begin{equation}
j_{i} (\nu,r,z) = \int_{m_e}^{M_\chi} dE \, \left(\td{n_{e^-}}{E}{} + \td{n_{e^+}}{E}{}\right) P_{i} (\nu,E,r,z) \; ,
\label{eq:emm}
\end{equation}
where $\td{n_{e^-}}{E}{}$ is the equilibrium electron distribution from DM annihilation (see \cite{gs2016,Colafrancesco2007} for details)
The flux density spectrum within a radius $r$ is then written as
\begin{equation}
S_{i} (\nu,z) = \int_0^r d^3r^{\prime} \, \frac{j_{i}(\nu,r^{\prime},z)}{4 \pi D_L^2} \; .
\label{eq:flux}
\end{equation}
In Coma we will assume the following halo data following~\cite{Colafrancesco2006}: the virial mass is given by $M_{vir} = 1.33 \times 10^{15}$ M$_{\odot}$, with virial concentration $c_{vir} = 10$, and that the density profile is of the Navarro-Frenk-White (NFW) form~\cite{nfw1996}. The thermal electron distribution and magnetic field profiles are taken from the best-fit values of \cite{briel1992} and \cite{bonafede2010} respectively. We also use an annihilation flux boosting factor from dense halo substructure in Coma that is $\sim 30$ following methods described in \cite{prada2013,ng2014}. We will show results both with and without this boosting factor.

Taking a spectral function $S_i$, we can compare it to data from the Coma cluster and Reticulum II and find $3\sigma$ confidence level exclusion limits on the value of $\langle \sigma V\rangle$ for the process $\chi \chi \to$ SM.

\section{Results and Discussion}
\label{sec:res}
Here we display the fraction of the cosmological DM abundance accounted for by Madala model DM. This is derived by determining the annihilation cross-section limits placed on the decay of Higgs-like coupled $S$, for a range of $\chi$ masses, by the spectra of the Coma galaxy cluster (radio and gamma~\cite{coma-radio2003,fermicoma2015}) and the Reticulum II dwarf galaxy gamma-ray spectrum~\cite{Fermidwarves2015}. If this cross-section falls below the range of the canonical thermal relic value ($2$ - $4$ $\times 10^{-26}$ cm$^3$ s$^{-1}$)~\cite{steigmann} then the ratio of the derived limit and lower end of the canonical band is taken to be the maximal fraction of DM accounted for. As the candidate model is already limited to annihilating too slowly to constitute all of the observed present epoch abundance.

\begin{figure}[htbp]
\centering
\includegraphics[scale=0.4]{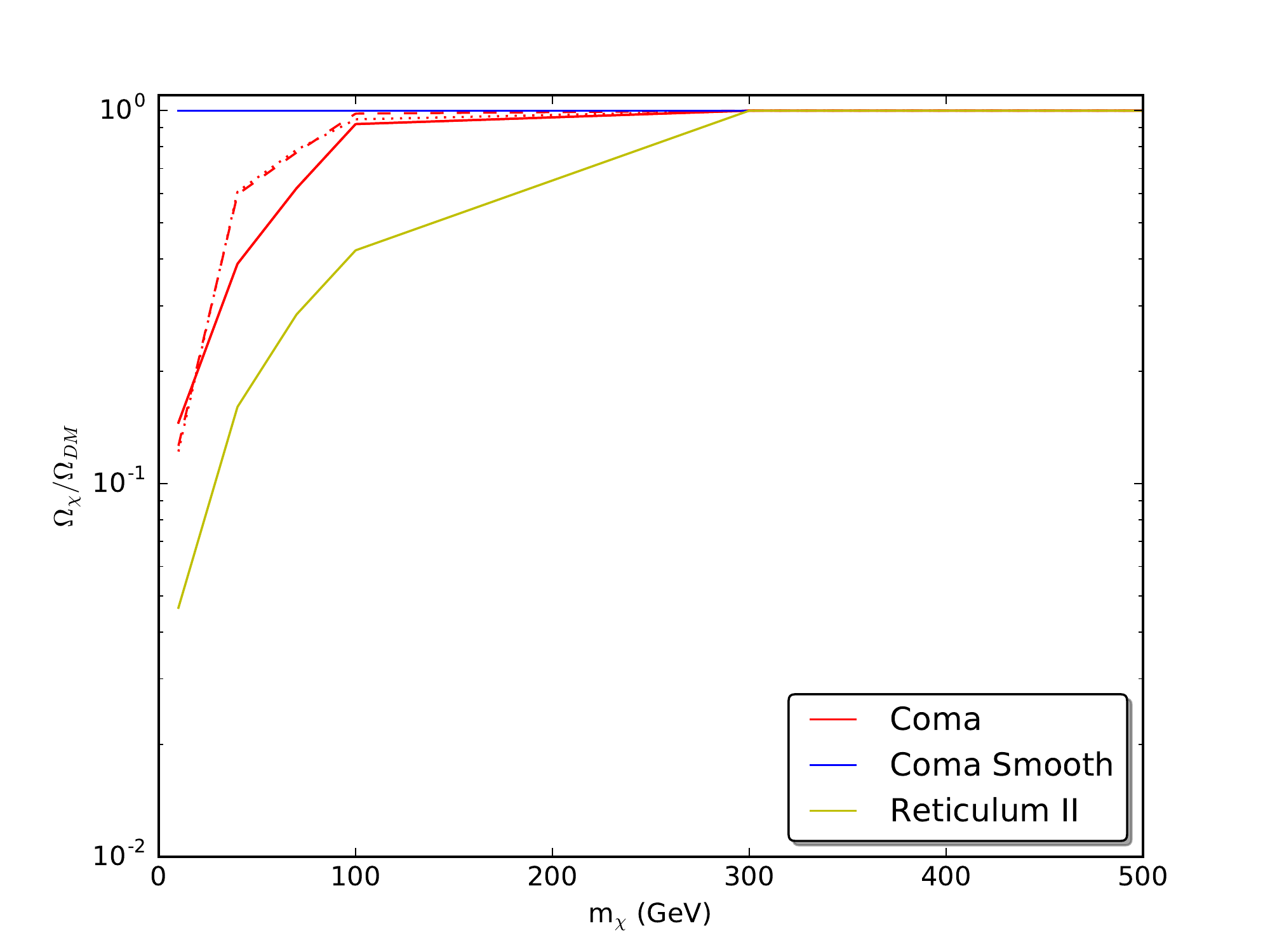}\includegraphics[scale=0.4]{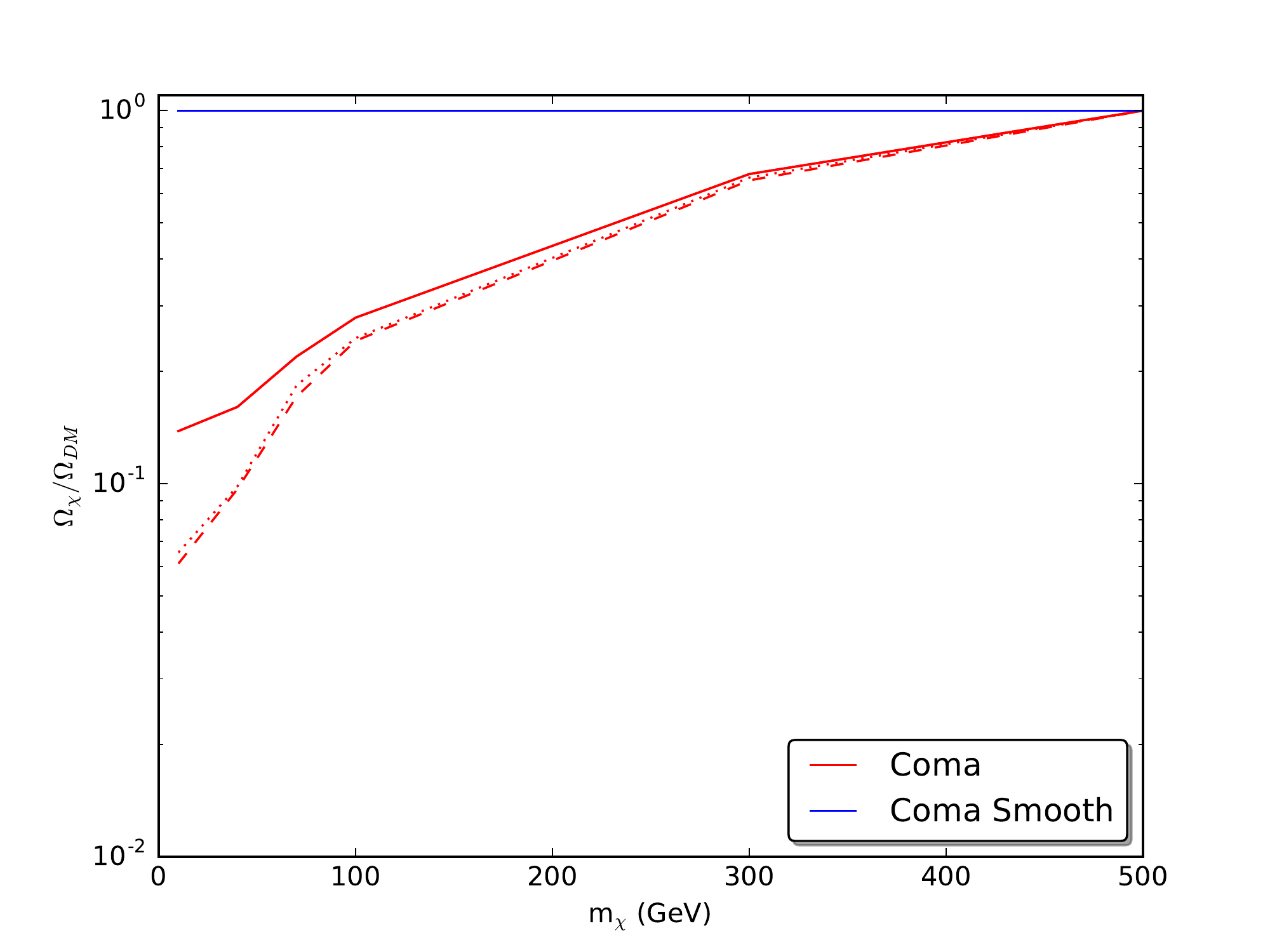}
\caption{Maximal cosmological DM fraction accounted for by Madala $\chi$ particles. The solid, dashed, and dotted lines correspond to $S$ masses 130, 160, and 200 GeV respectively. Coma Smooth shows the case without substructure boosting. Left: gamma-ray limits. Right: limits from Coma radio data.}
\label{fig:gamma}
\end{figure}

In the left-hand panel of Figure~\ref{fig:gamma} we show the DM fraction limits imposed on $\chi$ by the Fermi-LAT gamma-ray data. These indicate that Coma can only constrain this fraction for $m_{\chi} < 70$ GeV when the substructure boosting factor is used. However, the Reticulum II spectrum places strong limits that force $\chi$ to provide $< 40$\% of the DM abundance while $m_{\chi} \leq 100$ GeV. These limits become weaker as the $\chi$ mass increases because the upper-limits in Reticulum II rise as a power-law for higher frequencies and the peak of DM-induced spectrum shifts with $m_{\chi}$.

In the right-hand panel of Fig.~\ref{fig:gamma} we see that the radio limits are significant provided a boost factor of $\mathcal{O}(10)$ from substructure is assumed. Since this is conservative within the literature, and based on robust halo simulations~\cite{prada2013}, it is not an undue assumption in a structure as large as the Coma cluster. The limits in this case are very similar to those from Reticulum II. However, they become stronger for larger $m_S$, due to fact that the harder resulting spectra conflict with the spectral profile of diffuse radio emissions in Coma (due to increased $W$ boson production at larger $m_S$).

For a very general analysis of the Madala hypothesis DM candidate, under the same simplifying assumptions used to limit the space of free-parameters in the model~\cite{madala2} (that $S$ is Higgs-like), we have shown that current data from the Reticulum II dwarf galaxy and the Coma galaxy cluster is sufficient to limit the possible abundance of Madala-associated DM particles to $\mathcal{O}$(10\%) if $m_{\chi} \leq 100$ GeV. This expands on previous work~\cite{gsmadala} showing that it may be possible to use gamma-ray data to place limits on the couplings of $S$ to the standard model, and possibly even rule-out a Higgs-like $S$ if Madala DM is to constitute the entire cosmological abundance. Here we have also removed some of less general assumptions from the prior work (that only resonant $S$ production was considered).

Using the same modelling techniques as applied here, we will be able to generalise the analysis of \cite{gsmadala} to also allow the $\chi$ and $S$ masses to be independent, opening a much broader capacity to constrain the couplings of $S$ to the SM. This is in addition to the inclusion of radio data, which is shown to have similar independent constraining power to that of the previously employed gamma-ray data. We demonstrate this generalisation below in Figure~\ref{fig:gen}. This displays the $3\sigma$ confidence level limits on the branching ratio of $S$ to $W$ bosons from both gamma-rays (left panel) and radio data (right panel) when thermal relic annihilation cross-sections are assumed~\cite{steigmann}. This is displayed as a ratio of the branching ratio to that of $b_h$ the required fraction for a Higgs-like particle of mass $m_S$~\cite{smhiggs}. We see that the conclusion of \cite{gsmadala} is not substantially weakened by this generalisation, with evidence limiting the possibility of a Higgs-like $S$ boson coming from both Reticulum II gamma-ray data and Coma diffuse radio data while $m_{\chi} \lesssim 250$ GeV.

\begin{figure}[htbp]
\centering
\includegraphics[scale=0.4]{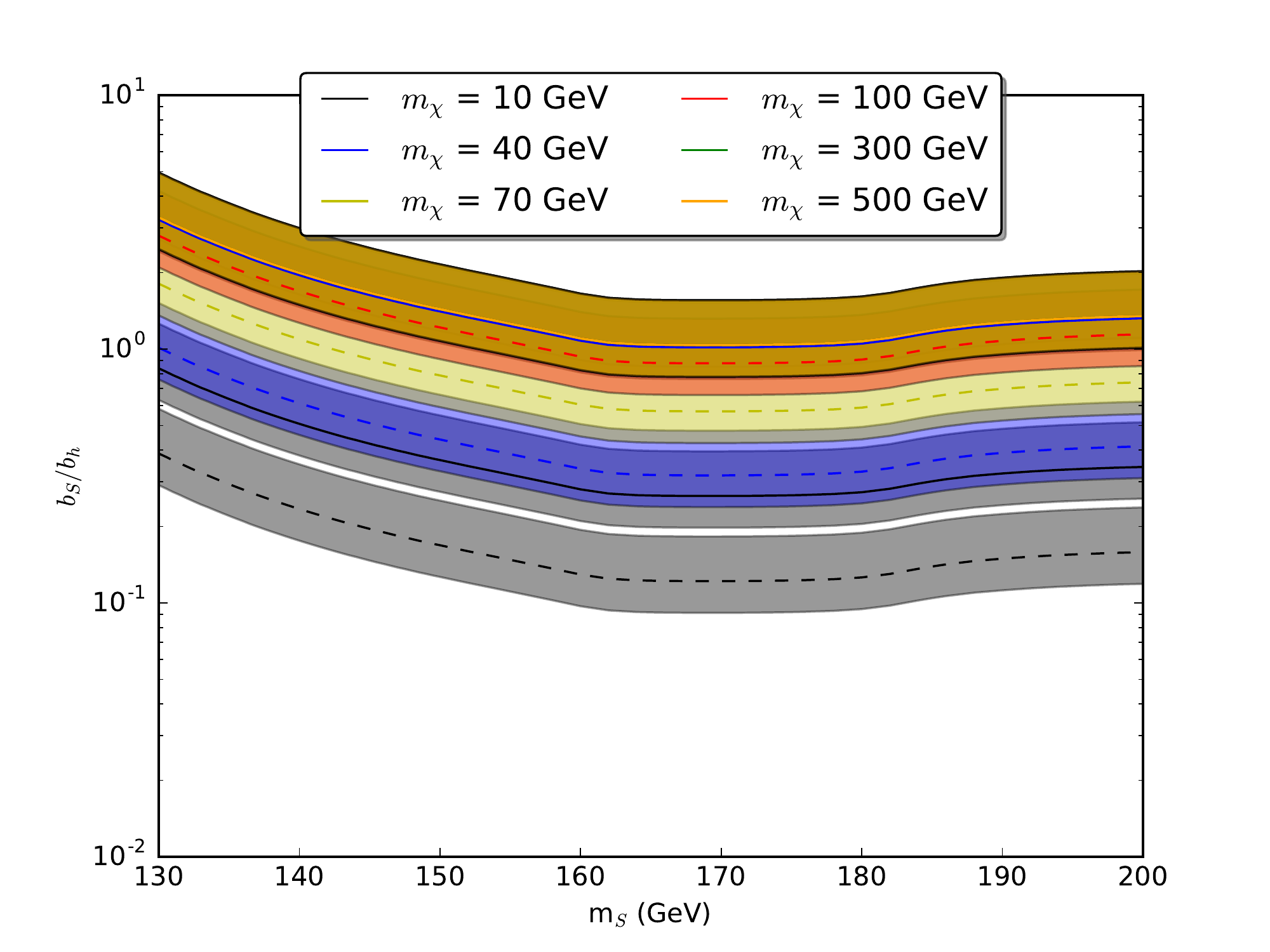}\includegraphics[scale=0.4]{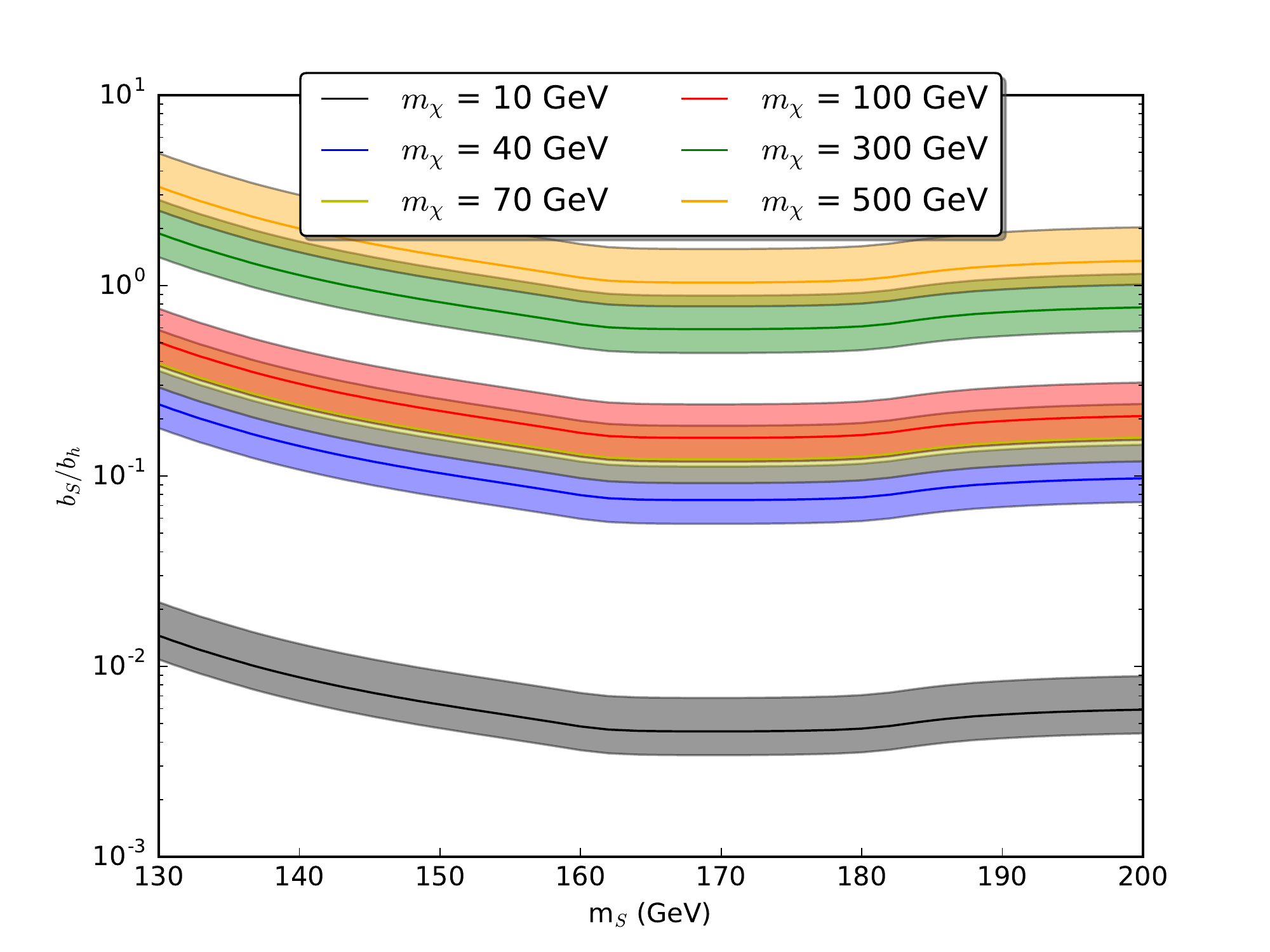}
\caption{Upper-limits at 95\% confidence level for the branching fraction into $W^+W^-$ compared to that required for Higgs-like couplings at a given $m_S$ when $\chi$ provides all DM. Shaded areas cover the thermal relic band region $2 - 4 \times 10^{-26}$ cm$^3$ s$^{-1}$~\cite{steigmann}. Left: gamma-ray limits with Coma in solid and Reticulum II in dashed. Right: limits from Coma radio data.}
\label{fig:gen}
\end{figure}

By bringing the viability of the simplifying Higgs-like assumption into question we can see that either many more free parameters must be added to the Madala model, which will require further astrophysical data to constrain accurately and may lower the significance of existing fits to excesses, or the $\chi$ particle of the Madala hypothesis must take a back seat as a candidate to explain all of DM. Both of these scenarios may well serve to weaken the attractiveness of the proposed Madala model. However, a final possibility is that the hidden sector particle $\chi$ is of large mass, $> 200$ GeV, limiting its ability to be put forward as an explanation in scenarios like the galactic centre gamma-ray excess~\cite{calore2014} as well as its detectability in collider experiments. This work remains especially relevant due to the persistence of the motivating LHC excesses into run-2 data~\cite{cern1}. 

\section*{Acknowledgments}
This work is based on the research supported by the South African Research Chairs Initiative of the Department of Science and Technology and National Research Foundation of South Africa (Grant No 77948).
G.B. acknowledges support from a post-doctoral grant through the same initiative and institutions.
We acknowledge useful discussions with B. Mellado on the Madala hypothesis proposed by members of the Wits-ATLAS group to account for several anomalies in both ATLAS and CMS data at the LHC.


\section*{References}

\begin{thebibliography}{99}

\bibitem{gsmadala}
G. Beck \& S. Colafrancesco, 2017, to appear in proceedings of Wits/iThemba labs High Energy Particle Physics Workshop 2017. arXiv: 1704.08031.

\bibitem{atlas-docs}
G. Aad et al., 2008, JINST, 3, S08003.

\bibitem{cms-docs}
M. Della Negra, A. Petrilli, A. Herve, \& L. Foa, 2008, \textit{CMS Physics Technical Design Report Volume I: Software and Detector Performance}, \url{http://doc.cern.ch//archive/electronic/cern/preprints/lhcc/public/lhcc-2006-001.pdf}

\bibitem{madala1}
S. von Buddenbrock, N. Chakrabarty, A. S. Cornell, D. Kar, M. Kumar, T. Mandal, B. Mellado, B. Mukhopadhyaya, \& R. G. Reed, 2015, preprint, arXiv:1506.00612 [hep-ph].

\bibitem{madala2}
S. von Buddenbrock, N. Chakrabarty, A. S. Cornell, D. Kar, M. Kumar, T. Mandal, B. Mellado, B. Mukhopadhyaya, R. G. Reed, \& X. Ruan, 2016, Eur. Phys. J. C, 76, 580. arXiv:1606.01674 [hep-ph].

\bibitem{madala3}
S. von Buddenbrock, 2017, arXiv: 1706.02477.

\bibitem{madala4}
Y. Fang, M. Kumar, B. Mellado, Y. Zhang, \& M. Zhu, 2017, Submitted to the White Paper of the Hong Kong University of Science and Technology Institute of Advanced Studies High Energy Physics Conference 2017. arXiv: 1706.06659.

\bibitem{cern1}
The ATLAS Collaboration, 2017, ATLAS-CONF-2017-058, \url{https://cds.cern.ch/record/2273874/files/ATLAS-CONF-2017-058.pdf}.

\bibitem{coma-radio2003}
M. Thierbach, U. Klein, \& R. Wielebinski, 2003, A\&A, 397, 53.

\bibitem{fermi-docs}
W. B. Atwood et al. for the Fermi-LAT collaboration, 2009, ApJ, 697, 1071. arXiv:0902.1089 [astro-ph]

\bibitem{fermicoma2015}
M. Ackermann et al for the Fermi-LAT collaboration, \& Y. Rephaeli, 2016, ApJ, 819 (2), 149. arXiv: 1507.08995 [astro-ph].

\bibitem{Fermidwarves2015}
A. Drlica-Wagner et al. for the Fermi-LAT collaboration \& T. Abbott et al. for the DES collaboration, 2015, ApJ, 809, L4. arXiv: 1503.02632 [astro-ph].

\bibitem{smhiggs}
S. Heinemeyer (ed) et al. for the LHC Higgs Cross Section Working Group, 2013, \textit{Handbook of LHC Higgs Cross Sections: 3. Higgs Properties : Report of the LHC Higgs Cross Section Working Group} (CERN). arXiv: 1307.1347 [hep-ph].

\bibitem{steigmann}
G. Steigman, B. Dasgupta, \& J. F. Beacom, 2012, Phys. Rev. D, 86, 023506.

\bibitem{pythia}
T. Sj\"ostrand, 1994, Comput. Phys. Commun., 82, 74..

\bibitem{darkSUSY}
P. Gondolo, J. Edsjo, P. Ullio, L. Bergstrom, M. Schelke, \& E.A. Baltz, 2004, JCAP, 0407, 008.

\bibitem{ppdmcb1}
M. Cirelli, G. Corcella, A. Hektor, G. Hütsi, M. Kadastik, P. Panci, M. Raidal, F. Sala, \& A. Strumia, 2011, JCAP, 1103, 051. Erratum: JCAP, 1210, E01 (2012). arXiv 1012.4515.

\bibitem{ppdmcb2}
P. Ciafaloni, D. Comelli, A. Riotto, F. Sala, A. Strumia, \& A. Urbano, 2011, JCAP, 1103, 019. arXiv 1009.0224.

\bibitem{micromegas1}
G. Belanger, F. Boudjema, A. Pukhov, \& A. Semenov, 2007, Comput. Phys. Commun., 176, 367. arXiv:hep-ph/0607059.

\bibitem{micromegas2}
G. Belanger, F. Boudjema, A. Pukhov, \& A. Semenov, arXiv:1305.0237 [hep-ph].

\bibitem{bonnivard2015}
V. Bonnivard, C. Combet, D. Maurin, A. Geringer-Sameth, S. M. Koushiappas, M. G. Walker, M. Mateo, E. Olszewski, \& J. I. Bailey III, 2015, ApJ, 808, L36.

\bibitem{gs2016}
G. Beck \& S. Colafrancesco, JCAP, 05, 013 (2016)

\bibitem{Colafrancesco2006}
S. Colafrancesco, S. Profumo, \& P. Ullio, 2006, A\&A, 445, 21.

\bibitem{Colafrancesco2007}
S. Colafrancesco, S. Profumo, \& P. Ullio, 2007, Phys. Rev. D, 75, 023513

\bibitem{nfw1996}
J. F. Navarro, C. S. Frenk, \& S. D. M. White, 1996, ApJ, 462, 563.

\bibitem{briel1992}
U. Briel et al., 1992, A\&A, 259, L31.

\bibitem{bonafede2010}
A. Bonafede, {\it et al.}, 2010, A\&A, 513, A3.

\bibitem{prada2013}
M. Sanchez-Conde \& F. Prada, 2014, MNRAS, 442 (3), 2271. arXiv:1312.1729 [astro-ph].

\bibitem{ng2014}
K. Ng et al., 2014, Phys. Rev. D, 89, 083001. arXiv: 1310.1915 [astro-ph.CO].

\bibitem{calore2014}
F. Calore, I. Cholis, C. McCabe, \& C. Weniger, 2015, Phys. Rev. D, 91, 063003. arXiv:1411.4647 [astro-ph]

\end{thebibliography}
\end{document}